\documentclass[aps,prb,twocolumn,groupedaddress,showpacs]{revtex4}

\usepackage{graphicx}
\usepackage[dvips]{color}
\usepackage{longtable}

\begin{document}


\title{Unusual Low-Temperature Phase in VO$_2$ Nanoparticles}




\author{Y. Ishiwata$^1$}
\author{S. Suehiro$^{1}$}
\author{M. Hagihala$^1$}
\author{X. G. Zheng$^1$}
\author{T. Kawae$^2$}
\author{O. Morimoto$^3$}
\author{Y. Tezuka$^4$}

\address{
$^1$Department of Physics, Saga University, Saga 840-8502, Japan \\
$^2$Department of Applied Quantum Physics, Kyushu University, Fukuoka 812-8581, Japan \\ 
$^3$Hiroshima Synchrotron Radiation Center, Hiroshima University, Higashi-Hiroshima 739-0046, Japan \\
$^4$Department of Advanced Physics, Hirosaki University, Hirosaki 036-8561, Japan
}


\date{\today}

\begin{abstract}
We present a systematic investigation of the crystal and electronic structure and the magnetic properties above and below the metal-insulator transition of ball-milled VO$_2$ nanoparticles and VO$_2$ microparticles.
For this research, we performed a Rietveld analysis of synchrotron radiation x-ray diffraction data, O $K$ x-ray absorption spectroscopy, V $L_3$ resonant inelastic x-ray scattering, and magnetic susceptibility measurements.
This study reveals an unusual low-temperature phase that involves the formation of an elongated and less-tilted V-V pair, a narrowed energy gap, and an induced paramagnetic contribution from the nanoparticles.
We show that the change in the crystal structure is consistent with the change in the electronic states around the Fermi level, which leads us to suggest that the Peierls mechanism contributes to the energy splitting of the $a_{1g}$ state.
Furthermore, we find that the high-temperature rutile structure of the nanoparticles is almost identical to that of the microparticles. 
\end{abstract}


\maketitle

\section{Introduction}

\begin{figure}
\includegraphics[width=8cm]{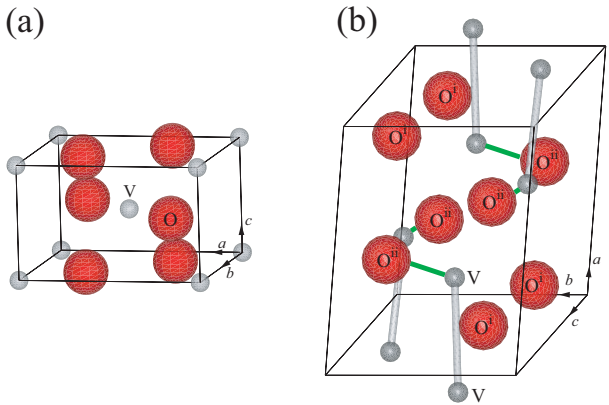}
\caption{(Color online) Crystal structure of VO$_2$ in (a) high-temperature rutile phase and (b) low-temperature monoclinic phase. 
V-V pairs and the shortest V-O bonds are illustrated.}
\label{f1}
\end{figure}

Since its discovery 50 years ago, the temperature-induced metal-insulator transition (MIT) in VO$_2$ has attracted significant attention.\cite{Morin1959,Goodenough1960,Adler1967,Goodenough1971,Zylbersztejn1975,Mott1990,Wentzcovitch1994,Rice1994,IFT1998,Tanaka2004,Biermann2005,Haverkort2005} 
In addition to structural and magnetic phase transitions near 340 K, VO$_2$ exhibits a resistivity jump of more than four orders of magnitude.\cite{Berglund1969,Mott1990,IFT1998}
The high-temperature metallic phase adopts a rutile ($R$) structure that involves the regular spacing of V$^{4+}$ ions along the $c$ axis, as shown in Fig. 1(a), and
the magnetic susceptibility for this phase is relatively large, which suggests the importance of electron correlations.\cite{Berglund1969}
The low-temperature insulating phase is a monoclinic ($M_1$) structure involving pairing and off-axis displacement of alternate V$^{4+}$ ions along the rutile $c_r$ axis, as shown in Fig. 1(b).
The formation of the V-V pair yields a nonmagnetic ground state.

\begin{figure}
\begin{center}
\includegraphics[width=8cm]{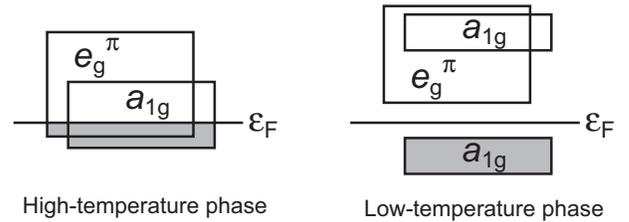}
\end{center}
\caption{Schematic energy diagram of V 3$d$ bands around the Fermi level for VO$_2$.\cite{Goodenough1971,Shin1990,Qazilbash2008}
}
\label{f2}
\end{figure}

Electronic states in VO$_2$ give rise to energy transfer near the Fermi level ($E_{\rm F}$) across the MIT, as shown in Fig. 2.\cite{Goodenough1971,Shin1990,Qazilbash2008}
The $a_{\rm{1}g}$ and the $e_{g}^{\pi}$ states originate from the crystal-field splitting of the V 3$d$ levels. 
The $a_{\rm{1}g}$ orbitals are directed along the $c_r$ axis and exhibit a nonbonding character in the high-temperature phase.
Furthermore, in the high-temperature phase, the $e_{g}^{\pi}$ orbitals mix with the O 2$p\ \pi$ orbitals, leading to a upward energy shift that results in their overlapping the energy of the $a_{1g}$ orbitals.  
The monoclinic distortion in the low-temperature phase causes the $a_{1g}$ orbitals to split into bonding and antibonding orbitals and the $e_{g}^{\pi}$ orbitals to split off from the lower $a_{1g}$ orbitals because of the strong $\pi$ bonding that results from the short V-O distance.
However, the role of V-V pairing versus electron correlations for the splitting of the $a_{1g}$ orbitals remains open to debate.\cite{Zylbersztejn1975,Wentzcovitch1994,Rice1994}

Recent experimental results have revealed a contribution of electron correlations to the MIT in VO$_2$.  
Infrared spectroscopy and nanoscale microscopy have shown divergent effective quasiparticle mass in approaching the insulating phase from the metallic phase in VO$_2$.\cite{Qazilbash2007} 
In addition, the MIT has been shown to be independent of the structural phase transition for thin films of VO$_2$ under an applied voltage,\cite{Kim2008} and the metallic monoclinic phase emerges under an applied pressure of over 10 GPa.\cite{Arcangeletti2007}   

In addition, by including the V-V pair as the cluster, the cluster-dynamical mean-field theory (CDMFT) yields the insulating $M_1$ phase in VO$_2$, 
leading to the proposition of a correlation-assisted Peierls-transition model in which both the electron correlations and the structural distortion contribute to the MIT.\cite{Biermann2005}  
Supporting these theoretical results is the fact that the spectral function obtained in the CDMFT has been experimentally confirmed using soft x-ray photoemission and O $K$ x-ray absorption spectroscopy (XAS).\cite{Koethe2006} 

Doping VO$_2$ with Cr creates the monoclinic $M_2$ and the triclinic $T$ phase, which indicates the importance of electron correlations in the MIT.\cite{Pouget1974} 
It is possible that the use of the nanostructures also give crystal structural changes to VO$_2$. 
Thin films of VO$_2$ have been shown to experience substrate strain, which decreases the band-gap energy in the low-temperature insulating phase.\cite{Kim1993} 
This result suggests that the nanostructures 
can be used to investigate the role of lattice distortion and V-V pairing in determining the spectrum of electronic states in VO$_2$.

Highly crystalline metal-oxide nanoparticles can be fabricated by grinding with a planetary ball mill.\cite{Zheng2005}
For VO$_2$ nanoparticles, the monoclinic distortion is expected to be affected by surface or finite-size effects, so it is important to elucidate the properties of VO$_2$ nanoparticles.
Therefore, we present herein a systematic investigation of the crystal and electronic structure and the magnetic properties above and below the MIT for ball-milled VO$_2$ nanoparticles and for VO$_2$ microparticles, the latter of which are used as a reference sample.
By performing a Rietveld analysis of synchrotron radiation x-ray diffraction (SR-XRD) measurements, we find an elongation and a reduction in the tilting of the V-V pair in the low-temperature monoclinic phase. 
In addition, O $K$ XAS and V $L_3$ resonant inelastic x-ray scattering (RIXS) measurements of the nanoparticles reveal that both the $e_{g}^{\pi}$ and the bonding $a_{\rm{1}g}$ state move to $E_{\rm F}$. 
Finally, we show that the change in the lattice distortion is consistent with the change in the electronic states around $E_{\rm F}$, 
which suggests that the Peierls mechanism contributes to the energy splitting of the $a_{1g}$ state.

\section{Experimental method}

\begin{figure}
\begin{center}
\includegraphics[width=8cm]{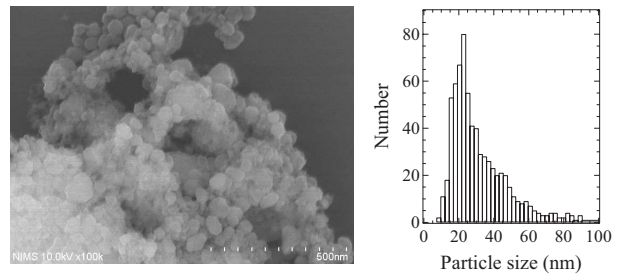}
\end{center}
\caption{(a) High-resolution SEM image and (b) statistical distribution of particle diameters for the VO$_2$ nano sample.}
\label{f3}
\end{figure}

VO$_2$ nanoparticles were prepared by ball milling from VO$_2$ microparticles (99\%, Furuuchi Chemical Corporation).
Both container and balls used were made of agate or  zirconia.
For experiments necessitating no surface oxidation, the mechanical milling was performed in H$_2$(1$\%$)/Ar gas; otherwise it was performed in air.
The air-milled nanoparticles were used in the SR-XRD experiments, whereas the H$_2$(1$\%$)/Ar-gas-milled nanoparticles were used in XAS, RIXS, and superconducting quantum interference device (SQUID) experiments. 
After ball milling, the XRD patterns acquired with an x-ray diffractometer (RINT-1100, Rigaku) that used Cu K$\alpha$ radiation were identical irrespective of atmosphere (data not shown).
Figure 3(a) shows a high-resolution scanning-electron-microscopy (SEM) image of the ball-milled particles and Figure 3(b) shows the statistical distribution of the particle diameters.
The histogram peaks at $\sim$25 nm and has a long tail that extends towards larger sizes.
The real variation in particle size should be narrower because it is difficult to find smaller particles due to their higher tendency to aggregate compared with larger isolated particles.  
The microparticles and the ball-milled nanoparticles are hereinafter referred to as ``bulk'' and ``nano'', respectively.

SR-XRD data were collected at the BL02B2 beamline of SPring-8 with 2$\theta$ ranging from 2.5$^{\circ}$ to 75$^{\circ}$ with 0.01$^{\circ}$ steps using a Debye-Scherrer camera with an imaging plate,\cite{Nishibori2001} and the incident x-ray wavelength was 0.50221 \AA.
All powder-diffraction data were analyzed by the Rietveld method using the program RIETAN-2000.\cite{Izumi2000}

XAS measurements in the partial-fluorescence-yield (PFY) mode and RIXS measurements were performed on the undulator beamline BL-2C at the KEK photon factory.\cite{Watanabe1997}
The soft x-ray spectrometer is based on the Rowland-circle geometry in which an input slit, a spherical grating, and a multichannel detector lie on the focal circle.\cite{Harada1998}  
The variable slit width was set to a maximum of 90 $\mu$m for the XAS experiment and a minimum of 10 $\mu$m for the RIXS experiment.
The multichannel detector was positioned to detect photons with energies from $\sim$485 to $\sim$545 eV.   
The resolution of the beamline spectrometer was set to about 0.1 eV, and the total resolution of the RIXS experiment was about 1 eV.
The incidence angle was $\sim$20$^{\circ}$ with respect to the sample surface. With respect to the incident beam, the spectrometer was placed at an angle of 90$^{\circ}$ measured in the vertical plane for the XAS measurements. For the RIXS measurements, it was positioned at the same angle, but measured in the horizontal plane.
Magnetic susceptibility measurements were performed using a commercial SQUID magnetometer at 5000 Oe.

\section{Results and Discussion}
\subsection{Synchrotron radiation x-ray diffraction and Rietveld refinement}

\begin{table*}
\caption{Refined structure parameters and reliability factors for the VO$_2$ bulk and nano samples. The crystal model for the refinement is space group $P$4$_2$/$mnm$, V in 2$a$, O in 4$f$ at 381 K; $P$2$_1$/$c$, V in 4$e$, O in two 4$e$ at 284 K.}
\begin{ruledtabular}
\begin{tabular}{ccccc}
	
			\    & Bulk (381 K) & Nano (381 K) & Bulk (284 K) & Nano (284 K) \\ 
		 \hline
			 $a$ [\AA] & 4.55072(3) & 4.5533(4) & 5.74613(10) & 5.755(3) \\
			 $b$ [\AA] & 4.55072(3) & 4.5533(4) & 4.52210(5) & 4.5289(11) \\
			 $c$ [\AA] & 2.85046(2) & 2.8586(3)& 5.37822(10) & 5.385(3) \\
			 $\beta$ [deg] & 90 & 90 & 122.6081(10) & 122.54(3) \\
			 $V$ [\AA$^3$] & 59.0303(7) & 59.267(9) & 117.723(4) & 118.32(8) \\
                   $x$ (V) & 0 & 0 & 0.23925(8) & 0.2377(3) \\
                   $y$ (V) & 0 & 0 & 0.97845(7) & 0.9817(4) \\
                   $z$ (V) & 0 & 0 & 0.02720(9) & 0.0179(4) \\
                   $B_{\rm{iso}}$ (V) [\AA$^2$] & 0.623(5) & 0.804(12) & 0.281(6) & 0.62(2) \\
                   $x$ (O$^{\rm{i}}$) & 0.29957(9) & 0.30025(16) & 0.1055(4) & 0.104(3) \\
                   $y$ (O$^{\rm{i}}$) & 0.29957(9) & 0.30025(16) & 0.2107(4) & 0.203(3) \\
                   $z$ (O$^{\rm{i}}$) & 0 & 0 & 0.2077(4) & 0.211(3) \\
                   $B_{\rm{iso}}$ (O$^{\rm{i}}$) [\AA$^2$] & 0.407(9) & 0.525 & 0.33(2) & 0.73 \\
                   $x$ (O$^{\rm{ii}}$) & - & - & 0.3985(4) & 0.405(3) \\
                   $y$ (O$^{\rm{ii}}$) & - & - & 0.7055(4) & 0.710 (3) \\
                   $z$ (O$^{\rm{ii}}$) & - & - & 0.2958(4) & 0.303(3) \\
                   $B_{\rm{iso}}$ (O$^{\rm{ii}}$) [\AA$^2$] & - & - & 0.33(2) & 0.73 \\
                   $R_{\rm{wp}}$ (\%) & 4.72 & 2.50 & 5.47 & 2.79 \\
			 $R_{\rm{I}}$ (\%) & 4.08 & 1.70 & 2.50 & 1.09 \\
                   $S$ & 2.22 & 1.39 & 2.55 & 1.55 \\

\end{tabular}
\label{Table.1}
\end{ruledtabular}
\end{table*}

\begin{figure}
\begin{center}
\includegraphics[width=8cm]{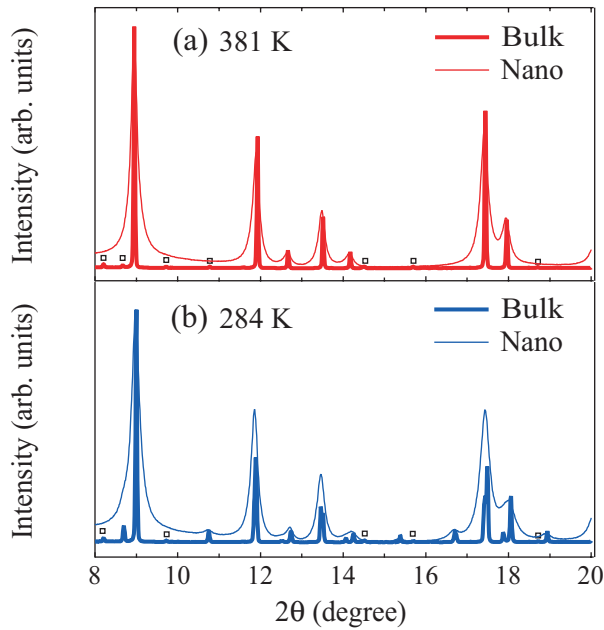}
\end{center}
\caption{(Color online) Low-angle region of SR-XRD patterns for the VO$_2$ bulk sample (thick solid line) and nano sample (thin solid line) at (a) 381 K and (b) 284 K. 
Some minor temperature-independent peaks are marked by open squares.}
\label{f4}
\end{figure}

Figure 4(a) shows the low-angle region of SR-XRD patterns for the bulk VO$_2$ samples (thick solid line) and nano VO$_2$ samples (thin solid line) at 381 K, which is above the MIT temperature.
The two XRD patterns are normalized to the respective maximum intensities.
Although the peak widths for the nano sample increase relative to the bulk sample (this is attributed to the reduction of the crystallite size, as clarified by a Rietveld analysis), the peak positions and the relative peak intensities of the two samples are almost constant.
These observations demonstrate that the crystal structure in the high-temperature phase does not change as the particle size is reduced by ball milling.
The XRD pattern for the bulk sample contains some minor peaks (open squares) that mostly correspond to the XRD pattern of V$_6$O$_{13}$.\cite{Bergstrom1998} 

Figure 4(b) shows the low-angle region of the SR-XRD patterns for the VO$_2$ bulk sample (thick solid line) and nano sample (thin solid line) at 284 K, which is below the MIT temperature.
The clear difference between the XRD patterns measured at 381 and 284 K is attributed to the structural phase transition.
The temperature-independent impurity peaks (open squares) are also present in the XRD pattern for the bulk sample at 284 K.

A comparison of the 284-K XRD patterns for the bulk and nano samples shows that the peak positions agree well, whereas the relative peak intensities differ significantly.
Thus, we conclude that the atomic positions in the low-temperature phase are displaced in the nano sample.

\begin{figure}
\begin{center}
\includegraphics[width=8cm]{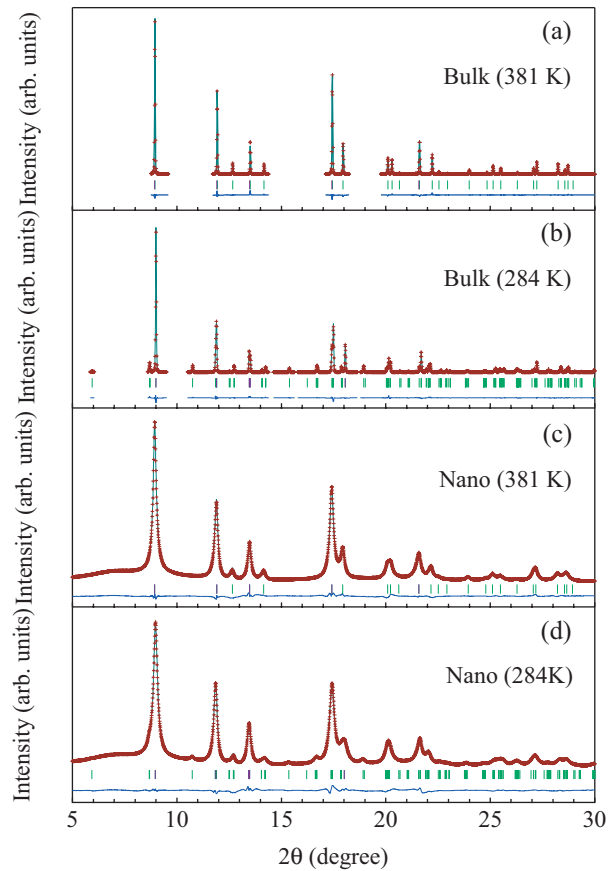}
\end{center}
\caption{(Color online) Observed and calculated SR-XRD patterns for the bulk sample at (a) 381 K and (b) 284 K and for the nano sample at (c) 381 K and (d) 284 K.}
\label{f5}
\end{figure}

\begin{table}
\caption{Bond distances and angles associated with V atoms in the low-temperature phases of the VO$_2$ bulk and nano samples.}
\begin{ruledtabular}			
\begin{tabular}{ccc}
	
			\     & Bulk (284 K) & Nano (284 K) \\ 
		 \hline
			 \shortstack{ \\ V-V [\AA] \\ (pair) } & \shortstack{ \ \\ 2.6109(7) \\ \ }  &  \shortstack{\ \\ 2.642(3) \\ \ }  \\
			 \shortstack{ V-V-V [deg] \\ (along the $c_r$ axis) } & \shortstack{ \ \\ 167.40(6) \\ \ }  &  \shortstack{\ \\ 170.71(12) \\ \ } \\
			 V-O$^{\rm{ii}}$ [\AA] & 1.739(2) & 1.792(12) \\
			 
\end{tabular}
\label{Table.2}
\end{ruledtabular}
\end{table}

The Rietveld analysis of the SR-XRD patterns for the VO$_2$ bulk and nano samples was performed using data in the range of $5^{\circ}< 2\theta< 74^{\circ}$, except for the region around the isolated impurity peaks for the bulk sample. 
We used the modified split pseudo-Voigt function in RIETAN-2000 as the peak profile function for all the XRD patterns.
Figures 5(a) to 5(d) show the result of the Rietveld fits for the VO$_2$ bulk and nano samples at 381 and 284 K.
For the high-temperature phase, the space group used for both the bulk and nano samples was $P$4$_2$/$mnm$ (tetragonal, rutile) with V$^{4+}$ ions occupying the 2$a$ site: 
$(0,0,0)$, $(\frac{1}{2},\frac{1}{2},\frac{1}{2})$ and O$^{2-}$ ions 4$f$: $(x,x,0)$, $(\bar{x},\bar{x},0)$, $(\bar{x}+\frac{1}{2},x+\frac{1}{2},\frac{1}{2})$, $(x+\frac{1}{2},\bar{x}+\frac{1}{2},\frac{1}{2})$.
For the low-temperature phase, the space group used for both samples was $P$2$_1$/$c$ (monoclinic) with V$^{4+}$ ions 4$e$:
$(x,y,z)$, $(\bar{x}, y+\frac{1}{2},\bar{z}+\frac{1}{2})$, $(\bar{x},\bar{y},\bar{z})$, $(x,\bar{y}+\frac{1}{2},z+\frac{1}{2})$ and O$^{2-}$ ions two 4$e$.
The ratio of the isotropic displacement parameters of the V atom to those of the O atom for the nano sample was fixed at the corresponding value obtained from the bulk sample.
The average crystallite size of the nano sample is estimated to be $\sim$18 nm by applying the pseudo-Voigt function of Thompson, Cox, and Hastings in the procedure of Finger {\it et al.} in RIETAN-2000.
The refined structure parameters and the reliability factors for the VO$_2$ bulk and nano samples are listed in Table I.
The structure parameters for the nano sample are one order of magnitude less accurate than those of the bulk sample.
For the high-temperature rutile phase at 381 K, the lattice constants and atomic coordinates of the bulk and nano samples differ by less than 0.3\%; so the nanoparticle effect is not clearly observed.
For the low-temperature monoclinic phase at 284 K, the lattice constants remain stable to within 0.2\%, whereas the atomic coordinates for V are found to vary significantly.

The bond lengths and angles associated with the V-V pair are listed in Table II [see also Fig. 1(b)].   
We find that the V-V-pair bond lengthens and that the alternating displacements of the V atoms perpendicular to the $c_r$ axis are reduced for the nano sample.
Furthermore, we find that the shortest V-O distance increases for the nano sample.

\subsection{X-ray absorption spectroscopy and resonant inelastic x-ray scattering spectroscopy}

\begin{figure}
\begin{center}
\includegraphics[width=8cm]{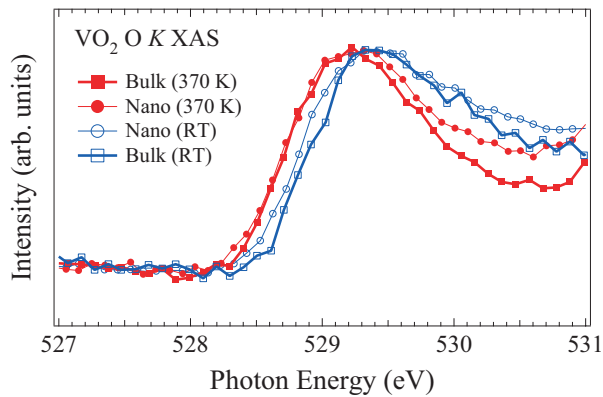}
\end{center}
\caption{(Color online) O $K$ XAS (PFY) spectra measured using a soft x-ray emission spectrometer for the VO$_2$ bulk sample (thick solid line with squares) and the nano sample (thin solid line with circles) at 370 K (solid symbols) and at room temperature (open symbols).}
\label{f6}
\end{figure}

Under O $K$ excitation, XAS is useful for investigating the unoccupied density of states on the V 3$d$ band because of covalent mixing between O 2$p$ and V 3$d$.\cite{deGroot1989}
Furthermore, taking advantage of the PFY mode using a soft x-ray emission spectrometer allows us to obtain a spectrum that is less sensitive to surface states.\cite{Ishiwata2010,Ishiwata2005}
Figure 6 shows O $K$ XAS spectra in PFY mode obtained for the bulk samples (thick solid line with squares) and nano samples (thin solid line with circles) at 370 K (solid symbols) and at room temperature (open symbols), which is above and below the MIT temperature.
The spectra are scaled to the height of the first peak, which is located at approximately 529 eV.
We find that, for the bulk sample, the O $K$ edge shifts by $\sim$0.2 eV between the two spectra measured above and below the MIT temperature, which is consistent with previous results for a single crystal of VO$_2$.\cite{Koethe2006} 

The high-temperature phase of the nano sample exhibits an absorption edge at an energy almost equal to that of the bulk sample, 
whereas the peak width increases toward the high-energy side of the first peak. 
The edge of the low-temperature phase of the nano sample redshifts about 0.1 eV relative to the bulk sample. 
The edge of the low-temperature phase is associated with the bottom of the $e_{g}^{\pi}$ states. 
These results show that the electronic states of the nano sample clearly change in going from the high-temperature to the low-temperature phase, which is similar to the behavior observed of the structure parameters, as discussed in the previous section.

\begin{figure}
\begin{center}
\includegraphics[width=8cm]{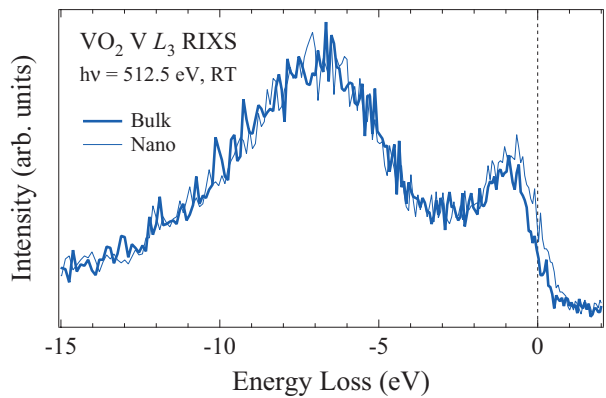}
\end{center}
\caption{(Color online) RIXS spectra under 512.5-eV excitation, which is near the V $L_3$ threshold for the VO$_2$ bulk sample (thick solid lines) and nano sample (thin solid lines) at room temperature.}
\label{f7}
\end{figure}

The x-ray photon energy ($h\nu_{\rm out}$) emitted by excitation of a core electron gives further information.
In particular, setting the incident photon energy ($h\nu_{\rm in}$) near the threshold of the metal 2$p$ state causes a coherent second-order optical process (i.e., absorption and emission of a photon), which renders accessible the $d$-$d$ transition that is dipole forbidden for a first-order optical process.\cite{Kotani2001}
Figure 7 shows the RIXS spectra plotted as a function of energy loss $h\nu_{\rm out} - h\nu_{\rm in}$ under 512.5-eV excitation near the V $L_3$ threshold for the VO$_2$ bulk samples (thick solid line) and nano samples (thin solid line) at room temperature (i.e., below the MIT temperature).
The two spectra are scaled to have the same intensity at both extremities in Fig. 7.
Based on previous results for a single crystal of VO$_2$,\cite{Braicovich2007} the features at $\sim$$-1$ and $-7$ eV are attributed to $d$-$d$ and charge-transfer excitations, respectively. 
The $d$-$d$ transitions of the nano sample have an energy loss of $\sim$0.3-eV less than that of the bulk sample.
The lowest $d$-$d$ transition in the low-temperature phase of VO$_2$ corresponds to the optical transition across the band gap from the lower $a_{\rm{1}g}$ level to the $e_{g}^{\pi}$ level. 
As a result, from the energy difference between the lowest $d$-$d$ transition and the O $K$ absorption edge for the nano sample, the upward shift of the $a_{\rm{1}g}$ state to $E_{\rm F}$ is estimated to be approximately 0.2 eV.  

\subsection{Magnetic susceptibility}

\begin{figure}
\begin{center}
\includegraphics[width=8cm]{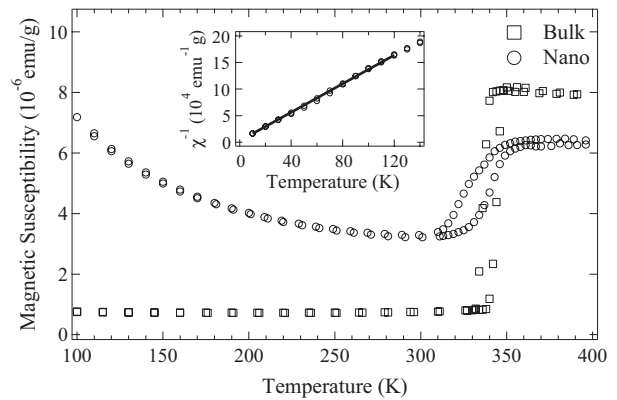}
\end{center}
\caption{Temperature dependence of magnetic susceptibility for VO$_2$ bulk sample (open squares) and nano sample (open circles). 
Inset shows the temperature dependence of the reciprocal magnetic susceptibility for the nano sample.
The solid line is a Curie's law fit.}
\label{f8}
\end{figure}

Figure 8 shows the temperature dependence of the magnetic susceptibility for the VO$_2$ bulk sample (open squares) and nano sample (open circles). 
The magnetic susceptibility of the bulk sample rises sharply with a hysteresis that is nearly constant because of the emergence of the nonmagnetic phase below the transition temperature. 
This result for the bulk sample is essentially in agreement with previous results.\cite{Berglund1969}
In contrast, the magnetic susceptibility for the nano sample monotonically increases with decreasing temperature in the low-temperature phase.
As shown in the inset of Fig. 8, the temperature dependence of the reciprocal of the magnetic susceptibility for the nano sample clearly obeys Curie's law.
The constituent ratio of the localized paramagnetic component is estimated to be about 16\% using the Curie constant, where we assume a spin $S = \frac{1}{2}$ with $g = 2$, which corresponds to a V$^{4+}$ ion.
Although it is possible that isolated V$^{4+}$ atoms exist near the surface,
the ratio of V surface atoms to the total number of atoms is approximately 3\% for 18-nm-diameter nanoparticles,\cite{ratio} which is equivalent to the average crystallite size in the nano sample.
This result suggests that some of the nanoparticles in the nano sample are paramagnetic even in the low-temperature phase.
However, further study using monodisperse nanoparticles is necessary to clarify this origin.

Above the transition temperature, the magnetic susceptibility of the nano sample is smaller than that of the bulk sample.
This result is not attributed to the addition of nonmagnetic impurities due to ball milling because the magnetic susceptibility is observed to be almost independent of milling time (data not shown).\cite{ms} 
In addition, nonmagnetic impurities are not observed in the SR-XRD patterns for the nano sample, as shown in Figs. 4(a) and 4(b). 
These results suggest that electron correlations are reduced in the high-temperature metallic phase of the nano sample.
The magnetic-susceptibility anomaly for the nano sample also appears in the width of the hysteresis.
The increase in the width of the hysteresis should be elucidated within the framework of classical nucleation theory, as was done in earlier work using nanoparticles buried in silica.\cite{Lopez2002}

\subsection{Discussion}
We have investigated the crystal and electronic structure and the magnetic properties of the VO$_2$ bulk and nano samples, and we find significant differences in the properties of the two samples in the low-temperature phases. 
The low-temperature phase of the nano sample involves the formation of elongated and less-tilted V-V pairs, a narrowed energy gap, and an induced paramagnetic contribution.

The O $K$ XAS of the low-temperature phase shows that the $e_{g}^{\pi}$ state of the lowest unoccupied state redshifts by approximately 0.1 eV for the nano sample relative to the bulk sample, which suggests that the ${\pi}$ bonding between the $e_{g}^{\pi}$ orbitals and the O 2$p$ orbitals becomes weak for the nano sample.
A qualitatively consistent result from the structural refinements is that the shortest V-O distance increases by about 3\% for the nano sample, as shown in Table II.
Furthermore, using the solid state table of Harrison\cite{Harrison1989} shows that the energy splitting due to the ${\pi}$ bonding decreases by about 0.25 eV for the nano sample relative to the bulk sample.\cite{pi}
The half of the estimated value is found to agree with the redshift observed in the O $K$ XAS.

For the low-temperature phase, the combination of the results of V $L_3$ RIXS and O $K$ XAS indicates that the bonding $a_{1g}$ state of the highest occupied state blueshifts by approximately 0.2 eV for the nano sample.  
If the system follows the Peierls model,\cite{Kagoshima1989} the energy splitting of the $a_{1g}$ orbital should be proportional to the displacement along the $c_r$ axis that results from the formation of the V-V pair.
The extent of the displacement is roughly determined by $(a_m/2 - l_{\rm{V-V}})/2$, where $a_m$ denotes the monoclinic $a$ axis (corresponding approximately to 2$c_r$) and $l_{\rm{V-V}}$ denotes the V-V bond length.
As a result, we find that, with respect to the values listed in Tables I and II, the displacement decreases by approximately 10\% for the nano sample.
The $a_{1g}$ splitting for the bulk sample is about 2.5 eV, which is equivalent to the results for single VO$_2$ crystals.\cite{Shin1990,Qazilbash2008} 
Assuming that electron-lattice interactions remain unchanged, we conclude that the $a_{1g}$ splitting decreases by about 0.25 eV for the nano sample, which is comparable to the shift of the bonding $a_{1g}$ state observed in the nano sample. This result suggests that the $a_{1g}$ splitting follows the Peierls model.
However, because recent experimental results show the crucial role of electron correlations in driving the MIT, \cite{Qazilbash2007} we interpret our results as indicating that the MIT is attributed to both effects; namely a correlation-assisted Peierls transition.\cite{Biermann2005}

The nano sample mainly consists of sphere-like nanoparticles, as shown in Fig. 1(a), which we attribute to the minimization of surface energy. 
Maintaining such a highly symmetric shape should require the suppression of lattice distortion, which could be attributed to the formation of the elongated and less-tilted V-V pairs for the nano sample.
In this case, controlling the particle size allows us to obtain further information, such as the relationship between the lattice distortion, the energy gap, and the paramagnetic contribution in the low-temperature phase. 


\section{Conclusion}
We observe the formation of an elongated and less-tilted V-V pair, a narrowed energy gap, and an induced paramagnetic contribution in the low-temperature phase of VO$_2$ nanoparticles by performing a Rietveld analysis of SR-XRD data, O $K$ XAS, V $L_3$ RIXS, and magnetic susceptibility measurements.
The change in the crystal structure is shown to be consistent with the change in the electronic states near $E_{\rm F}$. 
As a result, we suggest that the Peierls mechanism contributes to the energy splitting of the $a_{1g}$ orbital in this system.
The high-temperature rutile structure of the VO$_2$ nanoparticles is almost identical to the structure of the VO$_2$ microparticles, which exhibit bulk-like behavior. 
The change in the V-V pair within the nanoparticles is likely due to the competitive formation of a symmetric spherical shape.

\begin{acknowledgments}
We gratefully acknowledge the help of K. Kato and J. E. Kim at SPring-8 in the SR-XRD experiment.
We would also like to thank J. Adachi at KEK for assistance in the XAS and RIXS experiments, and we appreciate the support of Y. Inagaki at Kyushu University in carrying out the SQUID experiment.
We are furthermore grateful to D. Tsuya and E. Watanabe at NIMS for assistance in the SEM investigation.
This work was supported by the Takahashi Industrial and Economic Research Foundation.
SEM investigation was supported by the ``Nanotechnology Network Project" of the Ministry of Education, Culture, Sports, Science and Technology (MEXT), Japan.
\end{acknowledgments}


\newpage


\begin{references}
\bibitem{ratio} The ratio of V surface atoms to the total number of atoms is roughly determined by $\displaystyle \biggl(\pi d^2 / {\Bigl(\frac{V_{n,l}}{4}\Bigr)}^\frac{2}{3}\biggr) / \biggl(\frac{\pi d^3}{6} / \frac{V_{n,l}}{12}\biggr)$,
 where $d$ denotes the diameter of 18 nm and $V_{n,l}$ donotes the unit cell volume in the low-temperature phase of the nano sample that is obtained to be 118.32 \AA$^3$, as listed in Table I.
\bibitem{ms} The nano sample used in the magnetic susceptibility measurement in Fig. 8 was prepared by ball milling for 30 hours in H$_2$(1$\%$)/Ar gas.
Another sample milled for 50 hours in air, that was used in SR-XRD experiments, showed an almost identical result.
\end{references}
\end{document}